\documentclass[pre,aps,twocolumn,showpacs]{revtex4-1}

\usepackage{amsmath}
\usepackage{amssymb}
\usepackage{psfrag}
\usepackage{ulem}
\usepackage{graphicx}

\newcommand{\e}{\text{e}}

\newcommand{\ph}{\varphi}
\newcommand{\Ph}{\Phi}

\newcommand{\w}{\omega}

\newcommand{\tht}{\theta}

\newcommand{\vv}[1]{\mathbf{#1}}

\newcommand{\fig}[1]{\ref{fig:#1}}
\newcommand{\eq}[1]{(\ref{eq:#1})}
\newcommand{\SEC}{Sec.~}

\newcommand{\poinc}{Poincar\'{e} }
\newcommand{\roes}{R\"ossler }

\newcommand{\PRC}{\text{PRC}}

\begin{document}

\title{Optimal Phase Description of Chaotic Oscillators}


\author{Justus T. C. Schwabedal$^{1,2}$}
\email{jschwabedal@googlemail.com}
\author{ Arkady Pikovsky$^{2}$, 
Bj\"orn Kralemann$^{3}$, Michael Rosenblum$^{2}$}
\affiliation{Department of Physiology, Marburg University, 35037 Marburg,
Germany\\
$^{2}$Department of Physics and Astronomy, Potsdam University, 14476
Potsdam, Germany\\
$^{3}$Department of Education Science, Christian-Albrechts-University
Kiel, 24118 Kiel, Germany
}

\begin{abstract}
We introduce an optimal phase description of chaotic oscillations by
generalizing the concept of isochrones. On chaotic attractors
possessing a general phase description, we define the optimal isophases as
\poinc surfaces showing return times as constant as possible.
The dynamics of the resultant optimal phase is maximally decoupled of the
amplitude dynamics, and provides a proper description of phase resetting of
chaotic oscillations. The method is illustrated with the R\"ossler and Lorenz
systems.
\end{abstract}

\date{\today
}
\maketitle

\section{Introduction}

Phase description lies in the base of theory of self-sustained, autonomous
oscillators~\cite{Kuramoto1984,Pikovsky2001,Izhikevich-07}. 
A prudently defined phase
variable yields a one-dimensional description of the oscillator, allowing one
to characterize important aspects of its dynamics such as regularity of
oscillation, sensitivity to external forcing, etc. Moreover, the concept of
phase is important for data analysis of oscillatory processes in physics, chemistry, biology, and technical applications, where various
approaches exist for extracting different variants of phase variables
 from oscillatory scalar time series.

On a very basic level, every phase description starts with the
identification of those states of the oscillator which are in the same phase.
For a good phase description, the identification must be done in an invariant
way -- independent of the variables and observables used -- in order to make
statements about the oscillator's phase dynamics non-arbitrary and comparable.
The standard procedure of phase reduction is valid for periodic oscillators
that possess a stable limit cycle.  There, a certain family of \poinc sections,
called \textit{isochrones}, is used for the identification of states: Each
isochrone consists of those states which are mapped onto each other after one
oscillation period $T$, and which converge to the corresponding state on the
limit cycle \cite{Guckenheimer1975,Winfree1980}.

Even though chaotic oscillators do not possess a stable limit cycle, a
phase-like variable has been used for their description.  In this sense, the phase dynamics of chaotic
systems has been initially discussed in relation to diffusion properties
of phase~\cite{Farmer-81,Pikovsky-86b} and to phase synchronization
\cite{Pikovsky-85,Stone-92,Rosenblum1996}. However, to describe
these features, one does not need a good \textit{microscopic}, i.~e.~on the time
scale of the order of a characteristic period $T$, definition of the phase 
because both diffusion and synchronization are defined macroscopically, i.~e.~for
time scales much larger than $T$.
On the other hand, in the theoretical description of phase synchronization a
proper microscopic phase definition was pre-assumed
\cite{Rosenblum-Pikovsky-Kurths-97,Josic2001}, 
although no practical algorithm for construction of a 
phase variable with good
properties has been presented.
 The reason is that the chaotic phase
diffusion destroys a rigorous notion of time-coherent isochrones, because
any two states of the chaotic oscillator which 
are thought to be initially at the same
phase will diverge as their respective phases diffuse.

In this article, we suggest a numerical technique for phase description
of chaotic oscillations. For this goal, we construct special Poincar\'e
sections, which we call \textit{optimal isophases}. The choice of isophases 
is based on the properties of their return times.
In an application to chaotic oscillators, we
demonstrate an intimate relation between optimal isophases, chaotic phase
diffusion, and unstable periodic orbits. Specifically, we discuss the reduced
phase dynamics of chaotic oscillations and the decoupling of the amplitudes from
the phase dynamics. Next, we use the optimal phase to introduce a proper
framework for the description of phase resetting of chaotic oscillators.

Starting with an outline of the standard phase definition for periodic
oscillators via the isochrones, we introduce 
the generalized concept of isophases of chaotic oscillators in
\SEC\ref{sec:ipco}. In \SEC\ref{sec:dop}, certain dynamical properties
of the optimal phase are highlighted by the example of the \roes oscillator.
Thereafter, the relation between optimal isophases and unstable
periodic orbits is presented (\SEC\ref{sec:upo-isophase}). In \SEC\ref{sec:pls},
certain aspects of the theory are presented for the Lorenz oscillator.
In the last section we discuss our results.

\section{Isophases of periodic and chaotic oscillators} \label{sec:ipco}
\subsection{Periodic oscillators and their isochrones}
\label{sec:poti}

Phase is a natural variable for the description of periodic motions in
dynamical systems. It can be introduced in different ways, with different levels
of mathematical rigor
\cite{Guckenheimer1975,Winfree1980,Kuramoto1984,Pikovsky2001}. 
Here we outline
an approach that is mostly suited for a generalization to the case of chaotic
systems. 

The consideration starts with a general dissipative dynamical system showing
stable
periodic oscillations; the system's state $\vv{x}(t)$ is, thus, attracted to a
limit cycle $\vv{x}_0(t)$ having period $T$. In a vicinity of this periodic
attractor the state space can be foliated by a non-intersecting family of
\poinc sections $J(\ph)$, parametrized by a phase variable
$\ph$ with period
$2\pi$. 
With $J(\ph)$, a phase
variable $\ph(t)$ can be assigned to each
state of the trajectory $\vv{x}(t)\in J(\ph(t))$. Therefore, the family of
isochrones $J(\ph)$
provides a precise definition of what is meant by an oscillation: 
The system completes one \textit{oscillation} if the variable 
$\ph$ grows by
$2\pi$, i.e.~if the trajectory returns to a chosen isochrone, consequently
passing through all sections in
$J(\ph)$. In order to simplify nomenclature and to distinguish this variable
from the genuine phase, introduced below, we term $\ph$ as \textit{protophase}.
Introducing
coordinates on the sections $J(\ph)$, one can parametrize each point by a
vector of amplitudes $\vv{a}$ and the protophase $\ph$.

There are various equivalent ways to foliate the state space in such a way that
$\ph$
grows monotonically; for periodic oscillators with a period $T$, the optimal
foliation does exist \cite{Guckenheimer1975}. It can be introduced by
considering
the stroboscopic map $\vv{x}(t)\to\vv{x}(t+T)$. Clearly, all points on the
limit cycle are stable fixed points of this map. Hence, for each 
fixed point $\vv{x}_0$ there exist a stable manifold which
converges to $\vv{x}_0$ under the action of the stroboscopic map. These stable
manifolds, called \textit{isochrones}, constitute a special
foliation of the neighborhood of the limit cycle, for which by construction the
Poincar\'e map is the same as the stroboscopic map. 

In this way one introduces the phase of oscillation so that its time evolution
does not depend on the amplitudes $\vv{a}$. By virtue of a trivial
reparametrization $\varphi\to\theta=\frac{2\pi}{T}\int \frac{dt}{d\varphi}
d\varphi$  of this foliation, one can introduce the genuine phase
$\theta$ which grows strictly uniformly in time, with a constant instantaneous
frequency
$\dot\theta=\omega=2\pi/T$. This phase, defined in the whole basin of
attraction of the limit cycle, serves as a basis for a theoretical
description of perturbed periodic oscillations~\cite{Kuramoto1984}.  In
particular, one can easily formulate phase-resetting properties in terms of
this phase: If a state on the limit cycle $\vv{x}'$ is instantly perturbed to
some other state (even outside of the limit cycle), 
$\vv{x}'\to\vv{x}''$~, then the phase is reset by a value
$\Delta\theta=\theta(\vv{x}'')-\theta(\vv{x}')$~, which remains constant in
the course of further evolution (see also \SEC\ref{sec:prco} below).

Noteworthy, the extension of the phase to a vicinity of a periodic orbit 
can be defined either for a stable or unstable limit cycle. 
In the latter case, instead of using the stable manifold,
one constructs the isophases by using the unstable  manifolds of the fixed
points of the stroboscopic map. However, for saddle limit cycles having both
stable and unstable directions, this construction fails. Here one can construct
isochrones on the stable and unstable manifolds separately, but not in the whole
vicinity of the cycle.
With this in mind, we use below for the chaotic case, where isochrones
do not exist, the term ``isophases'' instead of the usual
``isochrones''.

\subsection{Protophase for chaotic oscillators}

We start the generalization of phase description to chaotic oscillators
by discussing the construction of the protophase. For this purpose, we need the
chaotic attractor to show the same property as a limit cycle, namely that there
exists a family of non-intersecting \poinc sections $J(\ph)$, monotonically
parametrized by a
protophase $\ph$. The requirement includes periodicity, $J(\ph+2\pi)=J(\ph)$,
and that any trajectory on the attractor successively crosses each \poinc
section $J(\ph)$
transversally. Of course, not all chaotic attractors possess such a family, but
those which have such a foliation can be described in terms of phases and are
the
subject of further consideration here. 

\begin{figure}[tbh]
\centering
\includegraphics[width=0.9\columnwidth]{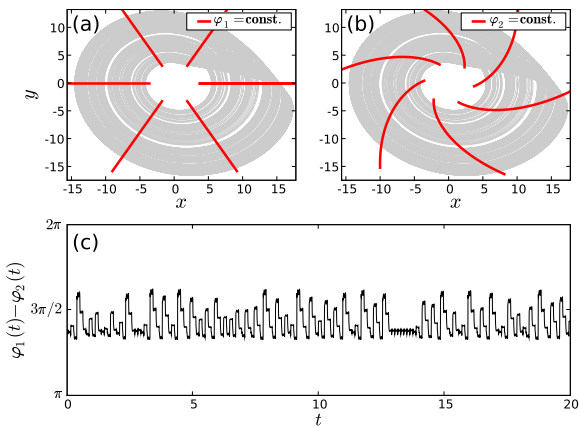}
\caption{ (Color online) 
(a,b) Two different families of \poinc sections or the \roes system
(red bold lines). Both families yield a proper definition of oscillation.
The corresponding protophases  $\ph_{1,2}$ are, however, different,
so that $\ph_2(t)-\ph_1(t)$ shows irregular bounded fluctuations (c), 
specific to the particular shapes of the \poinc surfaces. }
\label{fig:roes-intro}
\end{figure}

Let us consider as an example the \roes oscillator~\cite{Roessler1976}
\begin{equation}\begin{aligned}
	\dot{x} &=  -y-z~, \\
	\dot{y} &=  x + 0.15y~, \\
	\dot{z} &= 0.2+z(x-10)~,
\end{aligned}\label{eq:roes}\end{equation}
and take a family of \poinc sections
$J(\ph_1)$ defined via the cylindrical coordinates 
\begin{equation}
	\ph_1=\tan^{-1}\frac{y}{x}~;~~\vv{a}=(r,h)=\left(\sqrt{x^2+y^2},~
z\right)~.
	\label{eq:roes-cyl}
\end{equation}
This family of \poinc sections with constant protophase $\ph_1$ is shown in
Fig.~\fig{roes-intro}(a)). 
However, other families can be
defined as well; an example of another foliation based on the protophase
$\ph_2=\ph_1+0.7\ln r$ is counterposed in Fig.~\fig{roes-intro}(b). 

Because the difference of any two protophases is bounded, the asymptotic
properties of their phase dynamics, such as the mean frequency and the
diffusion constant of the phase rotations, do not depend on the definition of
the protophase. However, local, microscopic properties of the dynamics for two
protophases are different as it becomes apparent through the irregularly
fluctuating phase difference $\ph_1(t)-\ph_2(t)$ shown in
Fig.~\fig{roes-intro}(c). The fluctuations show a bounded and irregular pattern
that is specific to arbitrarily chosen variants of the \poinc sections. In
order to define a ``genuine'' phase, such as that of periodic oscillators, we
need to define the ``isophases'' of chaotic attractors. 
(Notice that for chaotic oscillators the isochrones generally do not exist.)
Because the phase of a
chaotic system is in fact not as ``genuine'' and unique as in the periodic case
(see discussion below), we will refer to it as \textit{optimal} phase, in the
sense that it represents oscillating properties of chaos in an optimal way.

\subsection{Optimal isophases for chaotic oscillators}
\label{sec:gico}

The genuine phase of periodic oscillators is defined by the basic property that
there exist \poinc sections where all return times are exactly equal to the
period
of oscillations; i.~e.~the corresponding Poincar\'e maps are stroboscopic maps
as well. Naturally, such a situation does not generally occur for chaotic
oscillators.
On the one hand, this is plausible because different periodic orbits embedded
in chaos usually have different basic periods (total period divided by the
number of crossings with a Poincar\'e surface, see
Eq.~(\ref{eq:upoProps2}) below). On the other hand, a
coincidence of Poincar\'e and stroboscopic map would also imply the absence of
phase diffusion what, however, is a degenerate, sparsely observed situation~\footnote{Such a coincidence one 
obtains for quasiperiodic regimes, where a smooth Poincar\'e section can be chosen as a stroboscopic 
map with one of the basic periods.}.

Since isophases of chaotic oscillators defined as sections with constant return
times do not exist in the strict sense, we introduce \textit{optimal isophases}
that approximate the property above with some accuracy. Practically, we
construct the optimal isophases as a smooth Poincar\'e section with a minimal
(bounded by the smoothness) variation of return times. As this condition is not
unambiguous, we describe below an algorithm that we practically use.

\begin{figure}[tbh]
\centering
\includegraphics[width=\columnwidth]{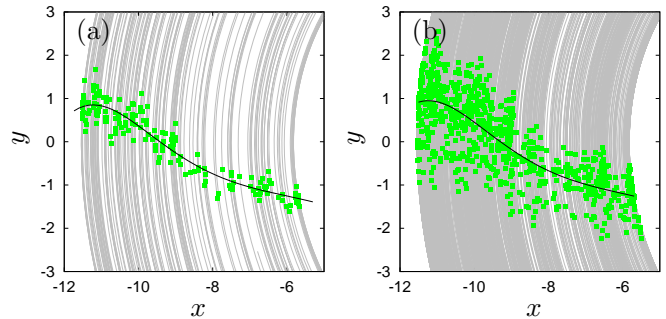}
\caption{ (Color online) The stroboscopic sets [Eq.~(\ref{eq:strset})] for the
R\"ossler attractor for two lengths of the trajectory  (a) $t_{end}=10^3$ and
(b) $t_{end}=5\cdot 10^3$ are shown by squares. The trajectories are shown with
gray lines, the optimal isophases obtained by fitting the
by a polynomial $\varphi(r)$ of order four are shown with black bold
lines.}
\label{fig:roes-str}
\end{figure}

The starting point of our construction is a suitable vector time series
$\vv{x}(t)$, $0\leq t\leq t_{end}$, of chaotic dynamics, which can be obtained by numerical simulation
or by embedding observed oscillations \cite{Kantz-Schreiber-04}. The first step is to
introduce an arbitrary 
protophase $\varphi$ as described above. Using it, we can estimate the average
period of oscillations as 
\[
T=\frac{2\pi t_{end}}{\varphi(t_{end})-\varphi(0)}\;.
\]
With this period, we define a family of \textit{stroboscopic sets} for the
trajectory
$\mathbf{x}(t)$ as
\begin{equation}
\mathbf{x}_k(\tilde\theta)=\mathbf{x}\left(\frac{\tilde\theta}{2\pi}
T+kT\right)\;,\quad k=1,2,\ldots,K_{end}\;.
\label{eq:strset}
\end{equation}
Here $\tilde\theta\in[0,2\pi)$ serves as a (still preliminary) phase
parametrizing stroboscopic sets, and each set consists of $K_{end}$ points.
These sets are invariant under the stroboscopic map with time interval $T$, 
but cannot serve as Poincar\'e maps as they are not smooth curves, because the
rotation in chaotic systems is non-uniform. The larger the total time interval
$t_{end}$, the stronger is the spreading of the points of the stroboscopic set. We
illustrate this in Fig.~\ref{fig:roes-str}. We note, that only in a degenerate
case where the phase diffusion of the chaotic oscillator vanishes, these
stroboscopic sets would be smooth lines which can be used as Poincar\'e
sections; such degenerate chaotic attractors (see an example
in~\cite{Rosenblum-Pikovsky-Kurths-97c}) possess the same rigorous phase
description as periodic oscillators.

In order to obtain a proper smooth Poincar\'e section, we fit the stroboscopic 
set, in the sense of least squares, by a polynomial
$\varphi=\varphi(\mathbf{a})$
(we use standard fitting procedure as described in~\cite{Press2002}). 
The resulting curves shown in Fig.~\ref{fig:roes-str} are our optimal isophases,
i.e. the curves of constant phase $\theta$. 

\begin{figure}[t]
	\centering
	\includegraphics[width=0.9\columnwidth]{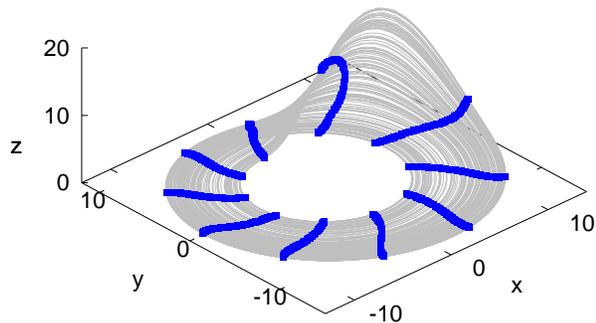}
	\caption{ (Color online) A global approximation of optimal isophases
			(blue dots which look like bold lines), obtained for the
\roes attractor (gray) using the approximation (\ref{eq:polyRep})
with $N_\ph=4$, $N_r=3$ and $N_h=1$. }
	\label{fig:roes-isoPlt2}
\end{figure}

If we restrict ourself to rather smooth isophases only, a good practical
approximation can be achieved if one introduces  a global phase correction
function $\Delta$ according to
\begin{equation}
\theta=\ph+\Delta(\ph,\vv{a})
\label{eq:coordTrafo}
\end{equation}
 and finds its representation in terms of
 polynomial basis functions: For each of the  amplitude
components $a_j$ we use the powers $a_j^n$, and for the phase variable $\ph$ we
use trigonometric polynomials $\exp( i\ph l)$. For example, for the R\"ossler
system in $1+2$
dimensions, consisting of phase $\ph$, radius $r$ and height $h$, the phase
correction is represented using a set of coefficients $c_{mnl}$:
\begin{equation}
	\Delta(\ph,r,h)=\sum_{m=0}^{N_r}\sum_{n=0}^{N_h}
	\sum_{l=0}^{N_\ph}c_{mnl}~r^m h^n\e^{il\ph}~.
	\label{eq:polyRep}
\end{equation}
The coefficients can be computed by applying  a linear least squares fit
\cite{Press2002} 
to the stroboscopic sets. In this way it is easy to find an
optimal phase globally, as a function of the state space coordinates
$\mathbf{x}$. 
We illustrate the isophases obtained in this way in Fig.~\ref{fig:roes-isoPlt2}.

\begin{figure}[t]
   \centering
   \includegraphics[width=0.6\columnwidth]{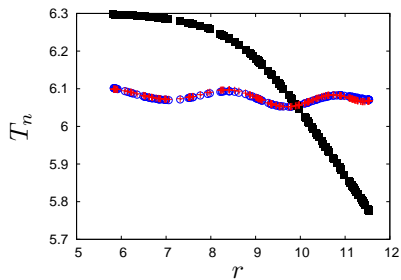}
   \caption{(Color online) Return times $T_n$ for the 
   \roes oscillator [Eq.~\eq{roes}]. Filled black squares correspond to an
arbitrary Poincar\'e section  $y=0,x<0$, here the spreading of the return times
is large. Local (blue open circles) and global (red crosses) approximations (nearly coinciding on the figure)
of the optimal isophases yield a strongly reduced spreading of the return times.
}
\label{fig:roes-arcLength2} 
\end{figure}

In Fig.~\ref{fig:roes-arcLength2} we compare the quality of the optimal
isophases obtained via representation (\ref{eq:polyRep}) with the results of the
local fitting of stroboscopic sets as in Fig.~\ref{fig:roes-str}. We compare the
return times for these isophases with the return times of the Poincar\'e section
$y=0,x<0$. One can see that globally defined
smooth isophases in the form (\ref{eq:polyRep}) give a quite good minimization 
of the variability of return times.

\section{Dynamics of the optimal phase}\label{sec:dop}
In this section we discuss dynamical properties of the optimal phase introduced with help of optimal isophases.
\subsection{Return time map}
\label{sec:rtm}
\begin{figure}[t]
	\begin{center}
		\includegraphics[width=0.9\columnwidth]{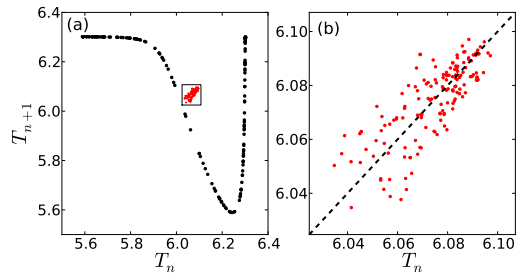}
	\end{center}
	\caption{ The return time map [Eq.~\eq{poincMap}] of the \roes oscillator
[Eq.~\eq{roes}] for the cylindrical \poinc section $\varphi=4\pi/3$ [Eq.~\eq{roes-cyl}] 
can be described as a one-dimensional chaotic
map (black dots in (a)).  Using the optimal isophase   
one obtains a map in a much smaller range (small box in (a) is enlarged in (b)). }
	\label{fig:roes-coher2}
\end{figure}

A natural way to characterize the time intervals $T_n$ between successive crossing of a Poincar\'e surface
is to construct the return time map
\begin{equation}
	T_{n+1}=M(T_n)~.
	\label{eq:poincMap}
\end{equation}
In fact, because $T_n$ is a function of the Poincar\'e map coordinate, it is just a scalar observable, and $M(T_n)$
is not a function but rather a one-dimensional projection of a Cantor set. Nevertheless, for  nearly two-dimensional 
strange attractors the Poincar\'e map is nearly one-dimensional, and (\ref{eq:poincMap}) looks like a curve 
(see Fig.~\ref{fig:roes-coher2}a). In Fig.~\ref{fig:roes-coher2} we demonstrate, how this return time map changes 
if one uses an optimal isophase as a Poincar\'e surface. First, the range of variations of $T_n$ drastically shrinks. 
Second, one can hardly recognize the one-dimensional structure of the map: because now $T_n$ is a ``bad'' observable,
it does not reproduce the nearly one-dimensional nature of the Poincar\'e map
$\vv{a}_n\to\vv{a}_{n+1}$.  This means, that with the dynamics of the new optimal phase 
looks like a random process even on a microscopic time scale, that is of the order of the period $T$.

\subsection{Uniformity of phase rotations}
\begin{figure}[t]
	\begin{center}
		\includegraphics[width=0.9\columnwidth]{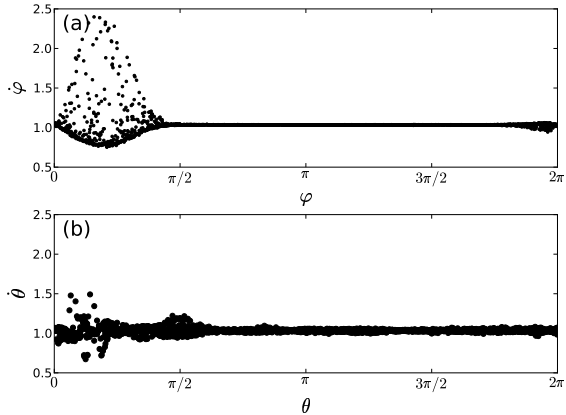}
	\end{center}
	\caption{ The phase velocities for the R\"ossler system 
for the protophase defined according to (\ref{eq:roes-cyl}) (panel (a)) and
according to isophases Fig.~\ref{fig:roes-isoPlt2}, panel (b). }
	\label{fig:roes-coher1}
\end{figure}

The basic property of the phase for a periodic oscillator is that it rotates uniformly. For the optimal phase of 
a chaotic oscillator we cannot expect pure uniformity, but nevertheless 
it should be considerably increased compared
to an arbitrary protophase. We illustrate this in Fig.~\ref{fig:roes-coher1}. 
Here we show the velocities of the protophase
$\ph$ defined according to Eq.~(\ref{eq:roes-cyl}) and that 
of the optimal phase $\theta$ defined according to isophases
shown in Fig.~\ref{fig:roes-isoPlt2}. While fluctuations in the protophase velocity $\dot{\ph}$
heavily depend on $\ph$, the fluctuations of $\dot{\tht}$ are almost uniformly
distributed and, notably, in some regions are larger than those of the
protophase.
Similar results are reported in Ref.~\cite{Kralemann2010}. We
conclude that optimal isophases not only eliminate the amplitude
dependence of the phase velocity, but they also flatten the phase dependence of its
velocity fluctuations.

\subsection{Decoupling of amplitude and phase dynamics}

\begin{figure}[t]
   \centering
   \includegraphics[width=0.9\columnwidth]{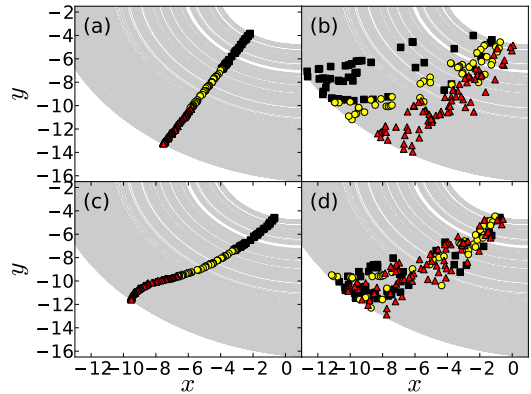}
   \caption{ (Color online) Two ``mixing'' setups where initial states of
the \roes oscillator [Eq.~\eq{roes}] (marked as symbols) are chosen either on 
the \poinc section $\phi_1=4\pi/3$ [Eq.~\eq{roes-cyl}]
(a) or on the optimal isophase (c).
Panels (b) and (d) show the same points at time $t=5\cdot T$, where $T$ is the 
average oscillation period. 
		States on the optimal isophase show less diffusive broadening in direction of the
   		phase than the points on arbitrary \poinc section. 
   		Moreover, states of different amplitude become indistinguishable only for the optimal
		isophase, as seen by the mixing of markers.  }
\label{fig:roes-info}
\end{figure}

One of the goals of introducing the phase is to decouple its dynamics from that of the amplitude.
For periodic oscillators this decoupling is perfect, whereas for chaotic
oscillators, it is only approximate.
To illustrate, how correlations of the phase dynamics with the amplitudes are reduced when the optimal
phase is introduced, we performed a ``mixing'' experiment; the results
are depicted in Fig.~\ref{fig:roes-info}.
We started an ensemble of initial conditions on a certain Poincar\'e surface and followed them for a time interval
of length $5T$ (five average rotation periods). The trajectories 
starting at small, medium, and large amplitudes $\vv{a}$ are marked separately in the plot.
In the upper panels (a,b), where a ``cylindrical'' Poincar\'e surface
$\phi_1=\text{const}$ is used, we see that the states
started at small amplitudes are lagged behind while those started at large amplitudes are advanced.
Contrary to this, using the optimal isophase as an initial condition, we see that
after 5 rotations all points are mixed and one can hardly 
distinguish the points which had different amplitudes at the beginning.
This is another illustration of the fact that the dynamics
of the optimal phase is effectively decoupled from the amplitude.

\subsection{Phase resetting of chaotic oscillators}
\label{sec:prco}

A basic application of the phase description of periodic oscillators is
quantification of the system response to pulse stimulation by means of
\textit{phase response curves}.
Given a state on the limit cycle $\vv{x}(\theta)$ one can determine the phase
shift due  to change of the state
$\vv{x}(\theta)\to\vv{x}'=\vv{x}(\theta)+\vv{k}$ simply by calculating
$\theta'=\theta(\vv{x}')$.  Because the phase rotates 
uniformly also outside of the limit cycle, the phase shift $\theta'-\theta$
remains invariant  and characterizes the phase 
resetting (for noise-induced  oscillations this notion can be also introduced in
an optimal sense~\cite{Schwabedal2010b}).  

\begin{figure}[tbh]
   \centering
   \includegraphics[width=0.6\columnwidth]{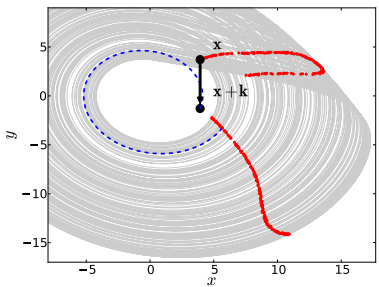}
   \caption{ (Color online)    
   A state $\vv{x}$ of the \roes attractor (gray trajectory) is kicked 
to the state $\vv{x}+\vv{k}$ (black points with an arrow). After one period
the perturbed trajectory (dashed line) returns to the attractor and now lies on the isophase $\theta[\hat{T}(\vv{x+k})]$ (red dotes). 
The kick's effect on the oscillator's phase is therefore given by the phase shift
$\PRC(\vv{k},\vv{x})=\theta[\hat{T}(\vv{x+k})]-\theta(\vv{x})$.   }\label{fig:roes-prc}
\end{figure}

\begin{figure}[t]
   \centering
   \includegraphics[width=0.9\columnwidth]{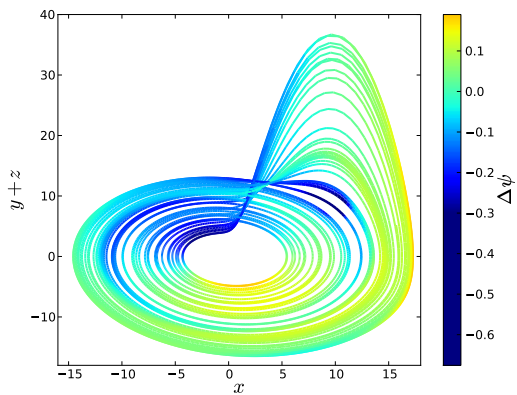}
   \caption{ (Color online) The phase resetting $\PRC(\vv{k},\vv{x})$  for the
states on 
the R\"ossler attractor, color (gray scale) coded, for $\vv{k}=(1,0,0)$.
   }
\label{fig:roes-prc-1}
\end{figure}

This 
approach has to be slightly modified when applied to chaotic oscillators. If both states $\vv{x}$ and $\vv{x+k}$ 
lie on the attractor, then their optimal 
phases are well-defined, and the phase shift can be simply calculated as  $\theta(\vv{x+k})-\theta(\vv{x})$. 
However, generally
the state  $\vv{x+k}$ lies outside of the attractor, and we have to generalize the definition of the optimal 
phases from the attractor to its vicinity. This is ambiguous, because the optimal isophases are not 
genuine isophases. They are not strictly invariant under time shifts  and we cannot define the phase of 
the state $\vv{x+k}$ by following its time evolution for arbitrarily large times. Instead, we have to fix the 
time interval after which the phase of the state  $\vv{x+k}$ is defined. For the R\"ossler model we choose
 the mean period $T$ as such an interval, as the relaxation time of approaching the attractor is typically 
smaller. So, we define
$\theta(\vv{x+k})=\theta[\hat{T}(\vv{x+k})]$ where $\hat{T}$ is the operator of time evolution over the 
average period $T$ (see Fig.~\fig{roes-prc}). Applying now the  representation (\ref{eq:polyRep}), we 
obtain the PRC  of the R\"ossler attractor $\PRC(\vv{k},\vv{x})=\theta[\hat{T}(\vv{x+k})]-\theta(\vv{x})$ 
as shown in Fig.~\ref{fig:roes-prc-1}.

\section{Isophases and Unstable Periodic Orbits} \label{sec:upo-isophase}

In this section we discuss a relation between optimal  isophases of a
chaotic system and \textit{unstable periodic orbits (UPOs)}
$\vv{x}_0(t+\tau)=\vv{x}_0(t)$ embedded in chaos.  For each UPO one can define
the phase on this orbit just from the condition of uniform rotation.  This
approach is discussed in \SEC\ref{sec:aops}. Similar to the construction
discussed in \SEC\ref{sec:poti}, we can extend the notion of the phase for
each periodic orbit to its stable or unstable manifold; these ideas are
presented in \SEC\ref{sec:oi}.

\subsection{Approximation of orbit phase sets}
\label{sec:aops}

For each UPO one can introduce a \textit{topological period} (lap number) $p$ as the number
of intersections with a Poincar\'e section. With this
number $p$ and the total period $\tau$, we define the \textit{oscillation period}
\begin{equation}
	S=\frac{2\pi}{\nu}=\frac{\tau}{p}~,
\label{eq:upoProps2} \end{equation}
which is expected to be close, but not identical, to the mean period of chaotic
oscillations (mean return time of
the Poincar\'e map).  Next, for the UPO we can introduce the phase $\tilde\theta$
that rotates uniformly with frequency $2\pi/\nu$ so that $\tilde\theta(\tau)=\tilde\theta(0)+2\pi p$. 
With help of this phase, a family of point sets $I(\tilde\theta)$, called
\textit{orbit
phase sets}, can be defined as points which are attained at constant time
intervals, equal to the oscillation period $S$:
\begin{equation}
	I(\tilde\theta)=\left\{\vv{x}_0(nS)~|~n=0,\dots,p-1\right\}~,
	\label{eq:pointSet}
\end{equation}
with some arbitrary choice of the zero phase.

\begin{figure}[t]
	\begin{center}
		\includegraphics[width=\columnwidth]{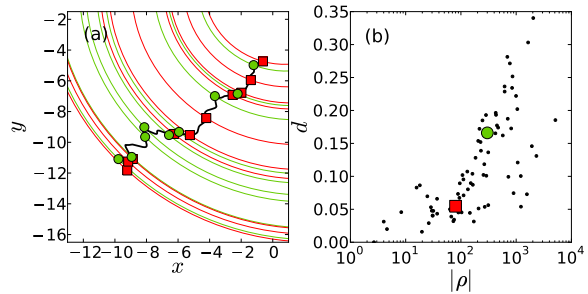}
	\end{center}
	\caption{ (Color online) (a): The optimal isophase
			(black line, obtained using a fit with a high-degree polynomial)
			of the \roes oscillator [Eq.~\eq{roes}]
			overlapped with orbit phase sets  [Eq.~\eq{pointSet}]
			 for a $9$-orbit (green circles) and a
			$10$-orbit (red squares).  (b): A distance measure
			$d$  [Eq.~\eq{upoDist}] quantifies how close is the
orbit phase set to the isophase, here shown for $80$ $p$-orbits with $p\leq10$.
One can see weak correlations to the instability of the UPOs measured by
the Floquet multiplier $|\rho|$. 
}
	\label{fig:roes-floq}
\end{figure}

Let us now take a Poincar\'e surface which passes through the orbit phase set
$I(\tilde\theta)$. (Of course, there are 
many possibilities to draw such a surface, e.g., one can use splines.)
Then it will be an approximation to an 
optimal isophase, as at least on the orbit phase set all the return times will
be equal to $S$. 
We illustrate this
with Fig.~\fig{roes-floq}(a), where we show orbit phase sets of two UPOs, with
topological periods  $p=10$ and $p=9$, for the R\"ossler
system~(\ref{eq:roes}). 
Since the orbits do not share any
state, the zero phases can be chosen separately. Practically, the phase offsets
have been chosen in a way that the orbit phase sets are  mostly close to each
other and approximate
the same isophase, which is also drawn for comparison.  
One can see that the orbit phase states indeed can serve 
as approximations for the isophases.

This approximation is expected to work better for larger 
periods and for periodic orbits which are most ``typical''.
  The
probability for a trajectory to approach the orbit depends on the
stability of the UPO, quantified by its unstable Floquet multiplier
\cite{Fairgrieve1991}. Therefore, it is expected that the correspondence between
isophases and
the orbit phase sets  will be better for UPOs which are visited more often
because they are less unstable.
To check this for the R\"ossler system,  we introduce a measure $d$ of the
distance of a $p$-orbit
$\vv{y}$ to the optimal isophase shown in Fig.~\ref{fig:roes-arcLength2} with a
green line, as
\begin{equation}
	d = \sqrt{p^{-1}\sum\limits_{k=0}^{p-1}||\vv{y}^J_k-\vv{y}_k||^2}\;,
	\label{eq:upoDist}
\end{equation}
where $\vv{y}^J_k$ are coordinates of the orbit phase set (for which we 
also optimized the zero phase to achieve a minimum of $d$)  and $\vv{y}_k$ are 
the crossings of the periodic orbits with the isophase.
This measure was calculated for the $80$ available
UPOs together with their Floquet multipliers. It was found that orbits showing
a larger distance had a tendency to be less stable
(cf.~Fig.~\fig{roes-floq}(b)).

\subsection{Orbit isophase}
\label{sec:oi}

\begin{figure}[t]
   \centering
   \includegraphics[width=0.9\columnwidth]{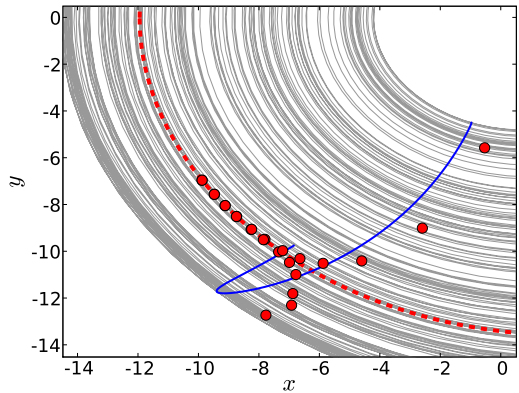}
   \caption{ (Color online) For a UPO of the \roes oscillator (dashed line), 
   the orbit phase set [Eq.~\eq{pointSet}] can
	be extended to the unstable manifold in two ways. Blue solid line shows the extension where the orbit's period $S$ is used.
	Red circles depict the extension based on the mean period of the chaotic attractor. The signularity of the latter curve indicates a divergence of the phase correction (see Appendix~\ref{appendix1} for the analytic form of this correction for the unstable Stuart-Landau oscillator.) 
		}
\label{fig:roes-Wu-1}
\end{figure}
As described in \SEC\ref{sec:poti}, after the phase on a periodic orbit is
introduced, the isophases in its vicinity can be defined separately on the
stable and unstable manifolds of the orbit, as the stable and the unstable
manifolds of the fixed points of the stroboscopic (with the period of the
orbit) map.  This definition can be applied to the UPOs in chaos, where the
unstable manifolds are especially interesting as they lie in the attractor.

Let us consider the simplest UPO of the \roes oscillator that has topological period
$p=1$. Its oscillation period is $S\approx6.024$ whereas the mean period of a
typical trajectory is $T\approx6.073$.  Numerically, we calculated the
isophase on the unstable manifold of this orbit using the oscillation period $S$
for the stroboscopic map and obtained the blue line in Fig.~\ref{fig:roes-Wu-1}. 
This isophase becomes folded together with the
unstable manifold, and is not close to the optimal isophases obtained by another methods.

It is instructive to try to construct the isophase on the unstable manifold of the UPO using not its period $S$,
but the mean period $T$. It is clear that such an isophase cannot exist, but trying approximate it
(see appendix~\ref{appendix1} for details) we obtain a singular curve  (Fig.~\ref{fig:roes-Wu-1}). 
This is another representation
of non-smoothness of stroboscopic sets that appears in the algorithm described in \SEC\ref{sec:gico}, due to non-existence
of true isophases. In fact, when one tries to construct an isophase, such a
singularity will appear for every periodic orbit,
and the procedure should be constrained by the requirement that the isophase
should be sufficiently smooth.

\section{Phase of the Lorenz System}\label{sec:pls}

\begin{figure}[tbh]
	\begin{center}
		\includegraphics[width=0.8\columnwidth]{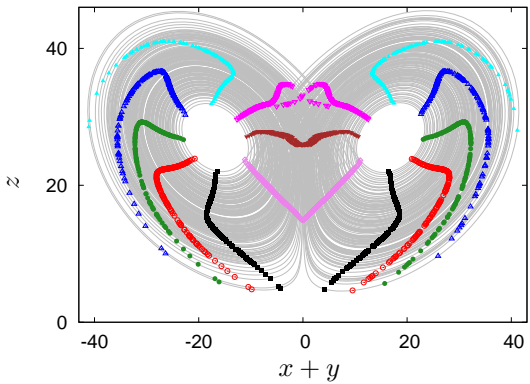}
	\end{center}
	\caption{ (Color online) Optimal isophases 
(different markers/colors depict different isophases) of the Lorenz attractor [Eq.~\eq{lorenz}] (grey line). 
		} 
\label{fig:lor-ip}
\end{figure}

In our presentation above we have used the R\"ossler model [Eq.~(\ref{eq:roes})] as the basic example. Here we 
discuss how the approach works for the Lorenz system
\begin{equation}
	\begin{aligned}
		\dot{x} &= 10\cdot(y-x)~, \\
		\dot{y} &= 28\cdot x-y-xz~, \\
		\dot{z} &= -\frac{8}{3}\cdot z+xy~.
	\end{aligned} \label{eq:lorenz} 
\end{equation}
Chaotic phase diffusion of the Lorenz system is orders of
magnitude stronger than that of the \roes oscillator Eq.~\eq{roes}, thus introducing its phase
is a more challenging task. The main difficulty lies in the unboundedness of the
return 
times of the Poincar\'e map, due to
presence of the saddle steady state at the origin ($x=y=z=0$). Due to this, the
stroboscopic sets 
are spread over the attractor and cannot serve as a basis for construction of
isophases like described above.  Therefore we applied the following iterative
procedure for obtaining smooth optimal isophases.
First, we use projections of the trajectory onto the plane $(u=\sqrt{x^2+y^2},~z)$.
On this plane the trajectory rotates around a center approximately at
$(12,27)$, and the protophases can be easily defined
(cf.~\cite{Pikovsky-Rosenblum-Osipov-Kurths-97}). We choose a Poincar\'e surface
and find the points of the trajectory at the intersection with this surface,
these are $x(t_k),y(t_k),z(t_k)$, $k=1,2,\ldots$. Of course, the times $t_k$ are
not equidistant because the Poincar\'e map is far from the stroboscopic one. We
adjust the times $t_k$ trying to make them equal, by introducing a parameter $s$
on which these times depend, and letting them evolve according to
\begin{equation}
\frac{dt_k}{ds}=-\frac{\partial V(t_1,t_2\ldots)}{\partial t_k}\;,\quad
V=\frac{1}{2}\sum_k(t_{k+1}-t_k-T)^2\;,
\label{eq:timeadj}
\end{equation}
where $T$ is the average period. 
One can easily see that the ``evolution'' of $t_k$ according to
Eq.~(\ref{eq:timeadj}) leads to equalization of the intervals $t_{k+1}-t_k$
because of minimization of the Lyapunov function $V$. However, we ``evolve'' the
times $t_k$ only for a finite interval of $s$, and obtain new times
$\tilde{t}_k=t_k(s)$. The new points $\tilde{x}
(\tilde t_k),\tilde{y}(\tilde t_k),\tilde{z}(\tilde t_k)$ form a new, 
distorted and singular Poincar\'e section. We smoothen this set by applying a
kernel technique~\cite{Hastie-Tibshirani-Friedman-01} and obtain a smooth
new Poincar\'e section with more equidistant time intervals. We make several
iterations of this procedure, and finally obtain the approximate smooth
isophases as depicted in Fig.~\ref{fig:lor-ip}.

\begin{figure}[tbh]
	\begin{center}
		\includegraphics[width=0.6\columnwidth]{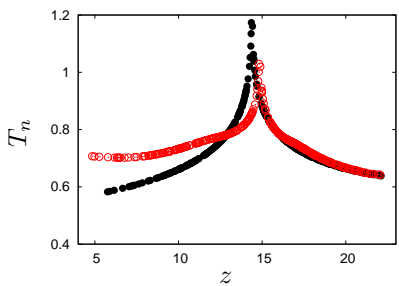}
	\end{center}
	\caption{ (Color online) Return times for the Poincar\'e section
$u=12,z<27$ (filled circles)  
and for the optimal isophases resulting from its iterations (shown in
Fig.~\ref{fig:lor-ip} with black filled squares) (open circles). The variations
of the return times only slightly decrease. 
		} 
\label{fig:lor-tim}
\end{figure}

To characterize the quality of the introduced isophases for the Lorenz system,
we plot the return times for an initial arbitrary Poincar\'e section and for the obtained isophase in 
Fig.~\ref{fig:lor-tim}. We see that the variations of the return times decrease only slightly, and 
the singularity (corresponding to the stable manifold of the origin) remains.

\begin{figure}[tbh]
	\begin{center}
		\includegraphics[width=0.9\columnwidth]{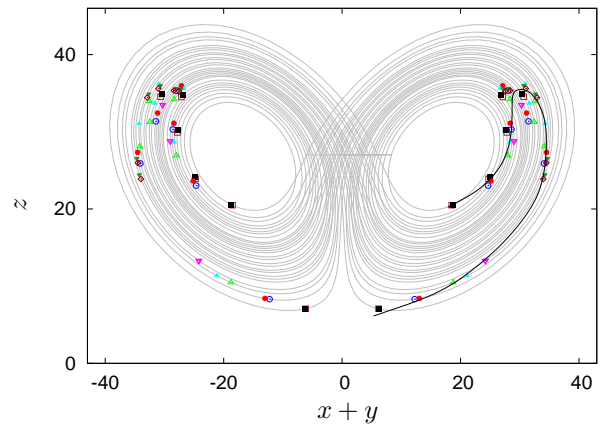}
	\end{center}
	\caption{ (Color online) Building an isophase using nine UPOs of the
Lorenz system with 
$p=6$ (see text for details).
		} 
\label{fig:lor-po}
\end{figure}

In Fig.~\ref{fig:lor-po} we use orbit phase sets of UPOs of the Lorenz system to
approximate isophases. 
Nine periodic orbits of the Lorenz system with topological period 6 are shown
with grey line. By manually
adjusting phase shifts of these orbits, it is possible to arrange the 
isophase sets (different markers) to build a set close to a curve (drawn
manually as a black line) which can serve as an optimal isophase. The form of
this curve is close to one of the optimal isophases presented in
Fig.~\ref{fig:lor-ip}.

\section{Conclusion}

In summary, we have proposed a method of phase reduction of chaotic oscillators by
generalizing the concept of standard isophases (isochrones). In the absence of a stable
limit cycle, the definition of optimal isophases of chaotic oscillations
is solely based on their return times. Because of non-vanishing chaotic
diffusion and embedded unstable periodic orbits with different periods,  isophases could
only be obtained in an optimal, approximate way constrained by certain smoothness
conditions. In the case of the R\"ossler attractor, where the phase diffusion is
relatively small, we obtain the optimal isophases by smoothing the stroboscopic
sets of a chaotic trajectory. For the Lorenz attractor, where phase diffusion is
large, an iterative numerical scheme was proposed. Using the 
\roes oscillator as an example, we have presented different aspects of the
phase dynamics. 
Specifically, the decoupling of the phase dynamics from the
amplitudes, as well as a way to describe phase resetting of chaotic oscillators
have been outlined.

The theory of optimal isophases can possibly provide a refined
understanding of emergent behavior of weakly coupled oscillating systems. For
example, a theoretical phase description of weakly coupled limit cycle
oscillators can be extended to ones of greater complexity, such as stochastic
or chaotic oscillators (cf.~\cite{Rosenblum-Pikovsky-Kurths-97,Josic2001}). 
In this way, one can treat more realistic models of natural systems.
Furthermore, the theory can easily be utilized in the analysis
of observed chaotic oscillations where the numerical scheme described above can be used to
refine a preliminary phase description. This can help to reduce certain
systematic errors which may be present in phase-related quantities such as
coupling strengths.

The theory is easily adaptable for the analysis of nonlinear oscillations with
a random component (for theoretical approaches see,
e.~g.~\cite{Yoshimura2008,Teramae2009,Goldobin2010}).  Here, the return times
to optimal isophases have to be interpreted in an average sense. The
corresponding results will be presented elsewhere.

\begin{acknowledgments}We thank M. Zaks for providing the data of UPOs of
R\"ossler and Lorenz systems, and R. Tonjes for discussions.
J.~S.~was supported by DFG 
(Collaborative Research Project 555 ``Complex Nonlinear Processes''), B.~K. 
was supported by Merz-Stiftung, Berlin.
\end{acknowledgments}

\appendix
\section{Isophases of unstable  Stuart-Landau oscillator}
\label{appendix1}

To give an analytically tractable example of isophases of UPOs on their unstable manifold, 
let us
consider the \textit{unstable Stuart-Landau oscillator} governed by
\begin{equation}
	\dot{r}=r(r^2-1);~\dot{\ph}=\alpha-\kappa r^2~.
	\label{eq:unstLS}
\end{equation}
It is exactly solvable: For the initial conditions $r(0)=R$ and $\ph(0)=\Ph$, it
has the well-known solution
\begin{eqnarray}
r(t) &=& \left[1+\frac{1-R^2}{R^2}\e^{2t}\right]^{-1/2}~, \label{eq:unstLS-polar1}\\
\ph(t) &=& (\alpha-\kappa)t-\kappa\ln r(t)+\Ph+\kappa\ln R~.
\label{eq:unstLS-polar}
\end{eqnarray}
Oscillator~\eq{unstLS} shows an unstable periodic orbit (UPO) with frequency
$\w=\alpha-\kappa$. Depending on the initial conditions, it either performs
decaying oscillations (for $R<1$), or diverges in finite time (for $R>1$).

As the characteristic period we first choose that of the UPO: $S=\frac{2\pi}{\w}$.
In order to obtain a phase which rotates independently of $r$, we set $\tht=\w
t+\Ph+\kappa\ln R$. Inserting $\tht$ into Eq.~\eq{unstLS-polar}, we
find that optimal isophases $I(\tht)$ are solutions of the equation
\begin{equation}
	\tht=\ph+\kappa\ln r~.
	\label{eq:isoEx1}
\end{equation}
For each $(\Ph,R)\in I(\tht)$ the return time for $\tht\to\tht+2\pi$ is equal
to $S$ because $\dot{\tht}=\w$. This is the standard definition of the isophases.

Alternatively, one may think of the unstable Stuart-Landau oscillator as of a
rarely visited part of the state space of a bigger chaotic system which has a
different
characteristic frequency $\frac{2\pi}{T}=\w_0=\w+\Delta\w$. It means, average
period $T$ is
different from the period $S$ of the UPO. 
Therefore, the condition that the return time for a Poincar\'e surface is equal to $T$ cannot
be fulfilled on the orbit. To fulfill the condition
for states off the periodic orbit, we
now seek a phase with the dynamics $\dot{\tht}=\w_0$. Therefore, we rewrite
Eq.~\eq{unstLS-polar} in terms of $\w_0t$:
\begin{equation}\begin{aligned}
	\ph(t) &= \w_0t+\Ph+\kappa\ln R-\kappa\ln r-\Delta\w t(r)~.\\
\end{aligned}\label{eq:isoEx2}\end{equation}
Here, we need to rewrite time as a function of radius, using (\ref{eq:unstLS-polar1}). We get
\begin{equation}
	t(r) = \frac{1}{2}\ln|r^2-1|-\ln r+\ln\frac{R}{\sqrt{1-R^2}}~.
	\label{eq:tofr}
\end{equation}
After the substitution, a uniformly rotating phase is given by
$\tht=\w_0 t+\Ph+\kappa\ln R+\Delta\w\ln(\sqrt{1-R^2}/R)$. Comparing the result
with Eq.~\eq{coordTrafo}, we obtain the phase correction as
\begin{equation}
	\Delta(r)=-(\kappa-\Delta\w)\ln r-\frac{\Delta\w}{2}\ln|r^2-1|~.
	\label{eq:isoEx3}
\end{equation}
While the return time is equal to $T$ off the periodic orbit, 
the phase correction diverges as $\ln|1-r|$ in the limit $r\to1$. Thus,
the ``isophase'' is singular and winds itself infinitely often around the limit cycle.

\end{document}